\documentclass[%
superscriptaddress,
twocolumn,
showpacs,
amsmath,amssymb,
aps,
prl,
floatfix,
showkeys,
reprint,
]{revtex4-1}

\usepackage{graphicx}
\usepackage{dcolumn}
\usepackage{bm}
\usepackage{amsmath}
\usepackage{amssymb}
\usepackage{multirow}
\usepackage{color}
\usepackage{booktabs}
\usepackage{tabularx}
\usepackage[english]{babel}
\usepackage[utf8]{inputenc}
\usepackage{enumerate}
\usepackage[cmyk,dvipsnames]{xcolor}

\usepackage[normalem]{ulem}

\begin{document}

\title{Optimal control of magnetization reversal in a monodomain particle by means of applied magnetic field}

\author{Grzegorz J. Kwiatkowski}
\affiliation{Science Institute of the University of Iceland, 107 Reykjav\'ik, Iceland}

\author{Mohammad H.A. Badarneh}
\affiliation{Science Institute of the University of Iceland, 107 Reykjav\'ik, Iceland}

\author{Dmitry V. Berkov}
\affiliation{General Numerics Research Lab, Moritz-von-Rohr-Stra{\ss}e 1A, 07745 Jena, Germany}

\author{Pavel F. Bessarab}
\email[Corresponding author: ]{bessarab@hi.is}
\affiliation{Science Institute of the University of Iceland, 107 Reykjav\'ik, Iceland}
\affiliation{ITMO University, 197101 St. Petersburg, Russia}
\affiliation{Peter Gr\"unberg Institute and Institute for Advanced Simulation, Forschungszentrum J\"ulich, 52425 J\"ulich, Germany}

\begin{abstract}
A complete analytical solution to the optimal reversal of a macrospin with easy-axis anisotropy is presented. Optimal control path minimizing the energy cost of the reversal is identified 
and used to derive time-dependent direction and amplitude of the optimal switching field. 
The minimum energy cost of the reversal 
scales inversely with the switching time for fast switching, follows an exponential asymptotics for slow switching and reaches the lower limit proportional to the energy barrier between the target states and to the damping parameter at infinitely long switching time. For a given switching time, the energy cost 
is never smaller than that for a free macrospin. This limitation can be bypassed by adding a hard anisotropy axis which activates the internal 
torque 
in the desired switching direction, thereby significantly reducing the energy cost. Comparison between the calculated optimal control path and minimum energy path reveals that optimal control does not translate to the minimization of the energy barrier, but signifies effective use of the system's internal dynamics to aid the desired magnetic transition.
\end{abstract}

\maketitle

Exact results concerning 
energy-efficient manipulation of 
magnetic 
structure are highly important for both fundamental science and technological applications as they could help improve performance of computing and memory devices based on magnetic elements. 
Optimization of magnetization switching in bistable nanomagnets by tuning the external magnetic field has come under special focus. 
It has been shown that a switching field can be significantly decreased by application of a weak radio frequency field pulse~\cite{thirion2003switching,woltersdorf_2007,zhu_2008,okamoto_2008,okamoto_2008a,bertotti_2009,wang_2009,yanes2009modeling,okamoto_2012,okamoto2014theory}. 
Magnetization reversal can be achieved exclusively by a microwave field~\cite{sun_2006}, whose amplitude can be reduced provided that the frequency is properly modulated~\cite{rivkin_2006,sun_2006a,cai2013reversal,klughertz_2014,islam_2018}. Sun and Wang~\cite{sun2006theoretical} obtained theoretical limit of the minimal switching field and derived an optimal constant-amplitude pulse yielding the shortest switching time. Barros {\it et al.}~\cite{barros2011optimal} developed a general theoretical framework for the design of control field pulses that minimize the energy cost of switching, calculated numerically the optimal switching field for a macrospin with easy-axis anisotropy and derived analytically the asymptotic properties of the reversal for infinitely long switching time~\cite{barros2013microwave}. 
So far, theoretical 
studies of optimal magnetization switching have been based on particular ansatzes for the switching field or involved numerical simulations, 
but a general analytical solution providing a transparent physical picture 
is still missing.   

Here, we present a complete analytical solution to the problem of energy-efficient switching of a nanomagnet with easy-axis anisotropy. 
Our results 
reveal new fundamental properties of the reversal including two asymptotic regimes of the energy cost 
and 
the optimal switching time. The easy-axis anisotropy can not reduce the energy cost of switching compared with the free-macrospin case, but this limitation can be lifted by introducing a hard anisotropy axis in the system.

To enhance efficiency of the magnetization reversal, it is important to minimize the energy losses associated with the generation of the switching field. Assuming an electric circuit to be the source of the field and neglecting the losses on radiation, the energy cost is defined by Joule heating due to the resistance of the circuit. This is proportional to the electric current square integrated over the desired switching time. Taking into account the linear relationship between the current magnitude and the strength of the generated field, we 
arrive at the cost functional in the form proposed by Barros {\it et al.}~\cite{barros2011optimal}: 
\begin{equation}
\label{eq:cost}
    \Phi = \int_0^T |\vec{b}(t)|^2dt,
\end{equation}
where $T$ is the switching time and $\vec{b}(t)$ is 
generated magnetic field at time $t$. 
The functional $\Phi$ 
needs to be minimized subject to specific boundary conditions and an equation of motion for the magnetic moment which is taken to be the Landau-Lifshitz-Gilbert equation,
\begin{equation}
\label{eq:llg}
    (1+\alpha^2)\dot{\vec{s}}=-\gamma \vec{s}\times(\vec{b}_i+\vec{b})-\alpha\gamma\vec{s}\times\bigl[\vec{s}\times(\vec{b}_i+\vec{b})\bigr],
\end{equation}
where $\alpha$ is the Gilbert damping, $\gamma$ is the gyromagnetic ratio, $\vec{s}$ is the unit vector along the magnetic moment $\vec{\mu}$ and $\vec{b}(t)$ is the external filed. The internal field is defined as $\vec{b}_\text{i}=-\mu^{-1}\partial E/\partial \vec{s}$ with $E$ being the internal energy of the system, i.e. the energy excluding the Zeeman term.

Constrained minimization of $\Phi$ can be formulated as an unconstrained optimization by expressing $\vec{b}$ in terms of the dynamical trajectory of the system as well as the internal magnetic field $\vec{b}_\text{i}$, 
\begin{equation}
\label{eq:bext}
    \vec{b}(t) = \frac{\alpha}{\gamma}\dot{\vec{s}}(t)+\frac{1}{\gamma}\left[\vec{s}(t)\times \dot{\vec{s}}(t)\right]-\vec{b}_\text{i}^{\perp}(t).
\end{equation}
Here, $\vec{b}_\text{i}^\perp\equiv\vec{b}_\text{i}-(\vec{s}\cdot\vec{b}_\text{i})\vec{s}$ is the transverse component of the internal field (the longitudinal component is not included as it does not affect the dynamics). 
Upon substituting (\ref{eq:bext}) into (\ref{eq:cost}), 
the energy cost of the reversal becomes a functional of the switching trajectory. By solving the Euler-Lagrange equation, the trajectory minimizing the cost functional $\Phi$ can be found. We denote this trajectory as the optimal control path (OCP) so as to distinguish it from other switching trajectories 
and to highlight its physical meaning. 
The optimal external field pulse can be obtained from the OCP using Eq.~(\ref{eq:bext}). 

\begin{figure}[!ht]
\centering
\includegraphics[width=1.0\columnwidth]{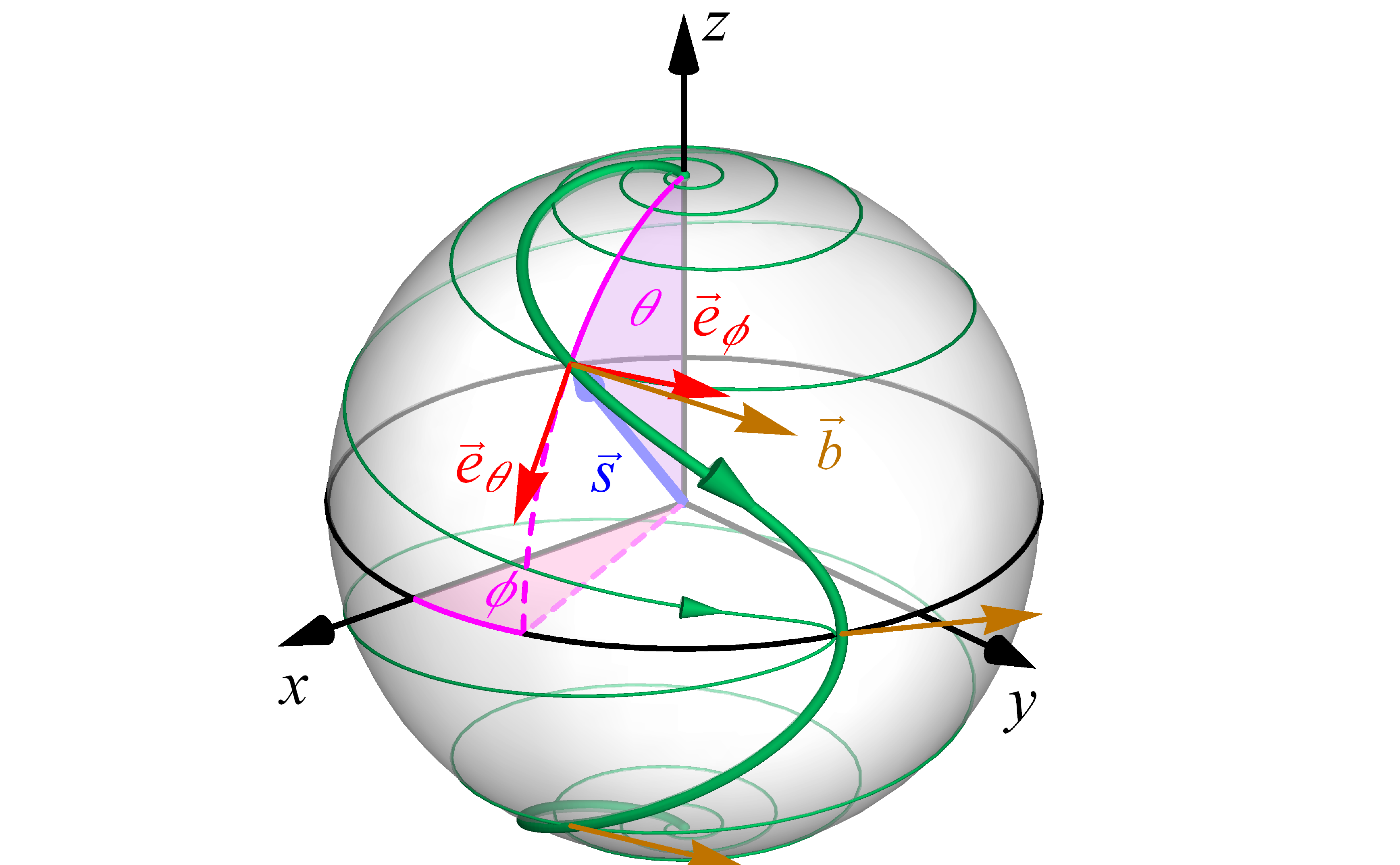}
\caption{\label{fig:system} Calculated optimal control paths (OCPs) for the reversal of a macrospin pointing along the unit vector $\vec{s}$. The initial and the final states are at the north and the south poles of the unit sphere, respectively. The damping factor $\alpha$ is $0.1$. The switching time $T$ is $10\tau_0$ and $100\tau_0$ for the paths shown with thick and thin green lines, respectively. 
External magnetic field $\vec{b}$ at $t=T/4$, $t=T/2$ and $t=3T/4$ is shown for the shorter path with the brown arrows.
}
\end{figure}

We apply the concept outlined above 
to the archetypal Stoner-Wohlfarth model, i.e. uniaxial monodomain particle whose magnetic moment is reversed by an external field (see Fig.~\ref{fig:system}). 
The internal energy $E$ of the system is defined by the anisotropy along $z$ axis,
\begin{equation}
    \label{eq:energy}
    E = -Ks_z^2,
\end{equation}
where $K>0$ is the anisotropy constant. 
Euler-Lagrange equations in spherical coordinates $\theta$ and $\phi$ (Fig.~\ref{fig:system}) 
read 
\begin{equation}
    \tau_0^2\ddot{\theta} =  \frac{\alpha^2}{4(1+\alpha^2)^2}\sin{4\theta}, \quad \tau_0\dot{\phi} =  \frac{\cos\theta}{1+\alpha^2}, \label{eq:ocp_tp}
\end{equation}
where the period of Larmor precession $\tau_0=\mu(2\gamma K)^{-1}$ defines the timescale. The boundary conditions $\theta(0)=0$, $\theta(T)=\pi$ correspond to the transition between the energy minima within the switching time $T$. 
Equation~(\ref{eq:ocp_tp}) for $\theta$ is the well known Sine-Gordon equation~\cite{mikeska1978solitons,cuevas2014sine} whose solutions are expressed by Jacobi elliptic functions: 
\begin{eqnarray}
    \theta & = & \frac{1}{2}\mathrm{am}\left(\left.\frac{t}{p\tau_0(1+\alpha^2)}\right|-\alpha^2 p^2\right), \label{eq:sol_t}\\
    \phi & = & \frac{1}{\tau_0}\int_0^{t} \frac{\cos(\theta(\tau))}{1+\alpha^2}d\tau+\phi_0, \label{eq:sol_p}
\end{eqnarray}
where $\mathrm{am}(.|.)$ is the Jacobi amplitude function \cite{supplement, abramowitz1948handbook}, $\phi_0$ is an arbitrary phase at $t=0$ and $p$ is a parameter implicitly defined through the following equation: $T = 4\tau_0(1+\alpha^2)p\mathcal{K}(-\alpha^2 p^2)$, with $\mathcal{K}(.)$ being the complete elliptic integral of the first kind \cite{supplement, abramowitz1948handbook}. 
The OCP described by Eqs.~(\ref{eq:sol_t})-(\ref{eq:sol_p}) reveals the mechanism for the magnetic moment reversal. 
The moment moves steadily from the initial state upward the energy surface while precessing counter-clockwise around the anisotropy axis until it reaches the top of the energy barrier at 
$t=T/2$. At this point, the precession reverses its direction and the system slides down to the target state minimum. 
This scenario was obtained numerically by Barros {\it et al.} \cite{barros2011optimal,barros2013microwave}, but the exact analytical solution makes it possible to derive general properties of the OCP~\cite{supplement}. 

Substitution of Eqs.~(\ref{eq:sol_t})-(\ref{eq:sol_p}) into Eq.~(\ref{eq:bext}) results in the following solution for the optimal switching field: 
\begin{eqnarray}
    \vec{b} & = & \frac{b}{\sqrt{1+\alpha^2}}(\alpha\vec{e}_\theta+\vec{e}_\phi),\label{eq:b_vec}\\
    b & = & \frac{K}{\mu p\sqrt{1+\alpha^2}}\left[\mathrm{dn}\left(\left.\frac{t}{p\tau_0(1+\alpha^2)}\right|-\alpha^2 p^2\right)\right.\nonumber \\
    & & \left.+\alpha p\ \mathrm{sn}\left(\left.\frac{t}{p\tau_0(1+\alpha^2)}\right|-\alpha^2 p^2\right)\right]\label{eq:b_amp}, 
\end{eqnarray}
where $\vec{e}_\theta$, $\vec{e}_\phi$ are local orthogonal unit vectors in the directions of increasing $\theta$, and $\phi$, respectively (see Fig.~\ref{fig:system}), while $\mathrm{dn}(.|.)$ and $\mathrm{sn}(.|.)$ are Jacobi elliptic functions \cite{supplement, abramowitz1948handbook}. 
Equation~(\ref{eq:b_vec}) reflects a general property of optimal switching protocols for axially-symmetric magnetic potentials~\cite{sun2006theoretical}. In particular, 
the switching field points in a specific fixed direction in the time-varying frame of reference associated with the magnetic moment. The orientation of the external field 
is such that its contribution to the precession around the anisotropy axis is exactly zero, which can be checked by substituting (\ref{eq:b_vec}) into (\ref{eq:llg}). 
Therefore, the external pulse contributes only to the part of motion which is relevant for switching, i.e. progressive 
increase in $\theta$. 
The optimal orientation of the switching field can be obtained regardless of optimization of the pulse amplitude, e.g. Eq.~(\ref{eq:b_vec}) still holds for the constant field amplitude, as demonstrated by Sun and Wang~\cite{sun2006theoretical}. 

\begin{figure}[!t]
\centering
\includegraphics[width=1.0\columnwidth]{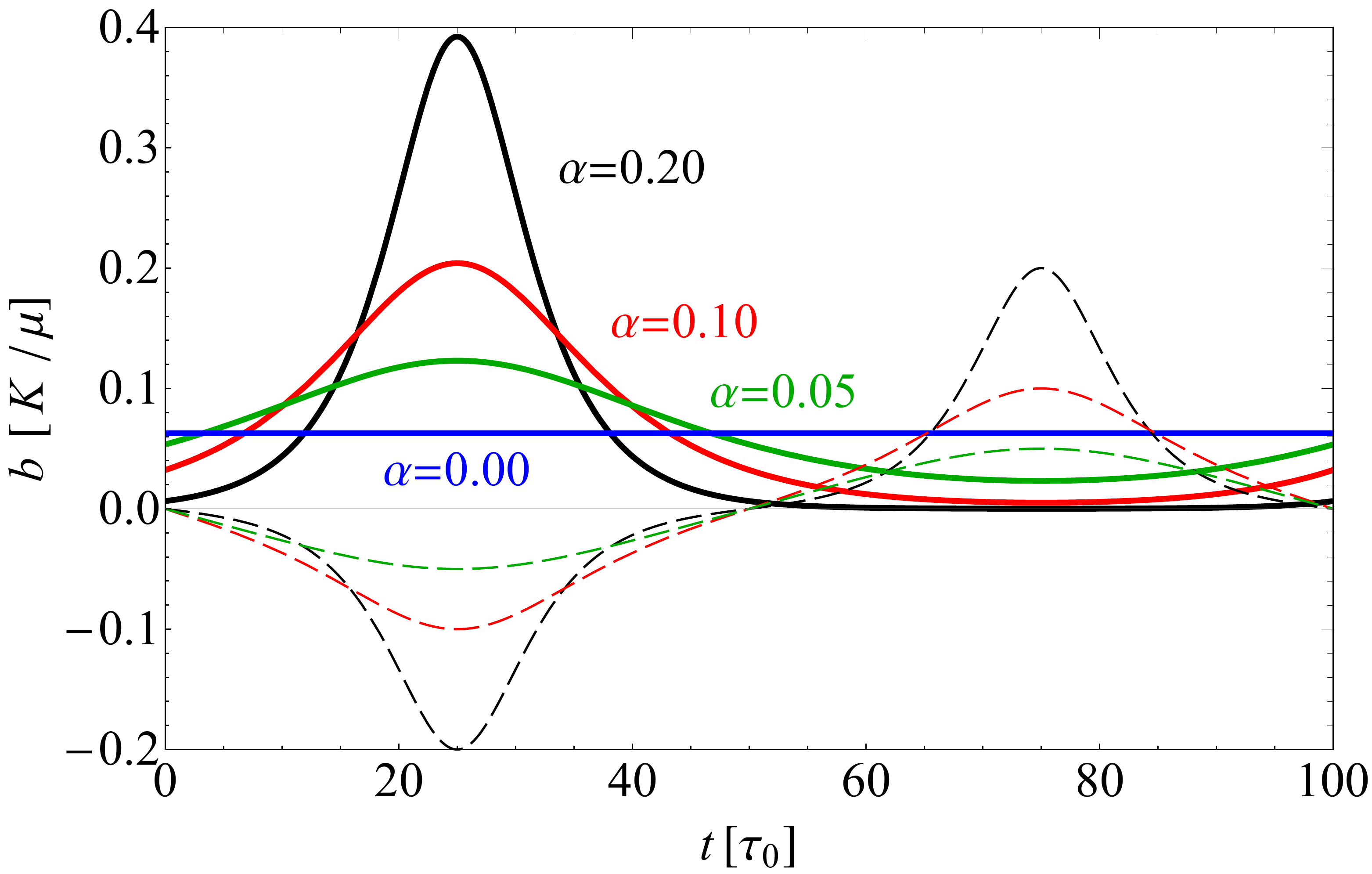}
\caption{\label{fig:pulse_t} Amplitude of the switching field as a function of time for $T=100\tau_0$ and several values of $\alpha$ (solid lines). 
Dashed lines show $\alpha b_\text{i}^{\perp}$ which is proportional to the polar component of the torque generated by the internal field. 
}
\end{figure}

Equation~(\ref{eq:b_amp}) describes the optimal switching field amplitude $b(t)$ (see Fig.~\ref{fig:pulse_t}). 
When $\alpha = 0$, the amplitude is time independent: $b(t)|_{\alpha=0}=\pi/(\gamma T)$. We emphasize here that for zero $\alpha$ there is no energy consumption by the magnetic moment itself. Nevertheless, the energy is still required to create the switching field which clearly demonstrates that the functional $\Phi$ does not describe energy dissipation by the magnetic system, but represents the energy spent by the external field source. 

For $\alpha > 0$, $b(t)$ has a more complex structure, but the symmetry  $b(0)=b(T/2)=b(T)$ holds. Damping gives rise to the internal torque in the polar direction. This torque - produced by the anisotropy field - counteracts the switching motion before crossing the equator, and a maximum in the switching field forms 
at $t=T/4$ so as to neutralize this effect (see Fig.~\ref{fig:pulse_t}). After the trajectory has crossed the equator at $t=T/2$, the internal torque aids the switching, and $b(t)$ reaches a minimum at $t=3T/4$. The position of the maximum and the minimum of $b(t)$ 
coincides with that of the extrema of the polar component of the internal torque (see Fig.~\ref{fig:pulse_t}). 
Note that the external field, although reduced compared to that before barrier crossing, is still non-zero in general: Some field needs to be applied in order to terminate the reversal on time. However, for long enough switching time, $T\gg (\alpha+1/\alpha)\tau_0$, damping alone is sufficient to complete the switching, and virtually no field needs to be applied after crossing the energy barrier (see black curve in Fig.~\ref{fig:pulse_t}). 
Although magnitude of neither maximum $b_{\text{max}}$ nor minimum $b_{\text{min}}$ of the switching field amplitude can be described in terms of elementary functions in a general case, the difference between them is always 
\begin{equation}
    \Delta b = b_\text{max}-b_\text{min}=\frac{2\alpha K}{\mu\sqrt{1+\alpha^2}}. \label{eq:b_delta}
\end{equation}
Moreover, the average amplitude $b_\text{av}$ can be computed analytically, leading to an exact relation
\begin{equation}
    b_\text{av} = \frac{1}{T}\int_0^T b(t)dt= \frac{\pi \sqrt{1+\alpha^2}}{\gamma T}, \label{eq:b_av}
\end{equation}
which demonstrates that overall larger fields are required in order to terminate the reversal in a shorter time, as expected. Interestingly, $b_\text{av}$ does not depend on the magnetic potential. From Eqs.~(\ref{eq:b_delta}), and~(\ref{eq:b_av}) it particularly follows that $\Delta b/b_{\text{av}}\rightarrow 0$ for $T\rightarrow 0$, i.e. decrease in the switching time progressively makes $b(t)$ resemble a time-independent function~\cite{supplement}. 

Equation~(\ref{eq:b_amp}) recovers the result of Barros {\it et al.} for $T\rightarrow \infty$ (see Eq. (13) in \cite{barros2013microwave}) as well as that of Sun and Wang for $\alpha=0$ (see Eqs. (7) and (9) in Ref.~\cite{sun2006theoretical}).
Additionally, for $T\ll (\alpha+1/\alpha)\tau_0$ the pulse amplitude simplifies to $b(t)\approx b_\text{av}+\Delta b \sin{\left(2\pi t/T\right)}/2$. 
\begin{figure}[!t]
\centering
\includegraphics[width=1.0\columnwidth]{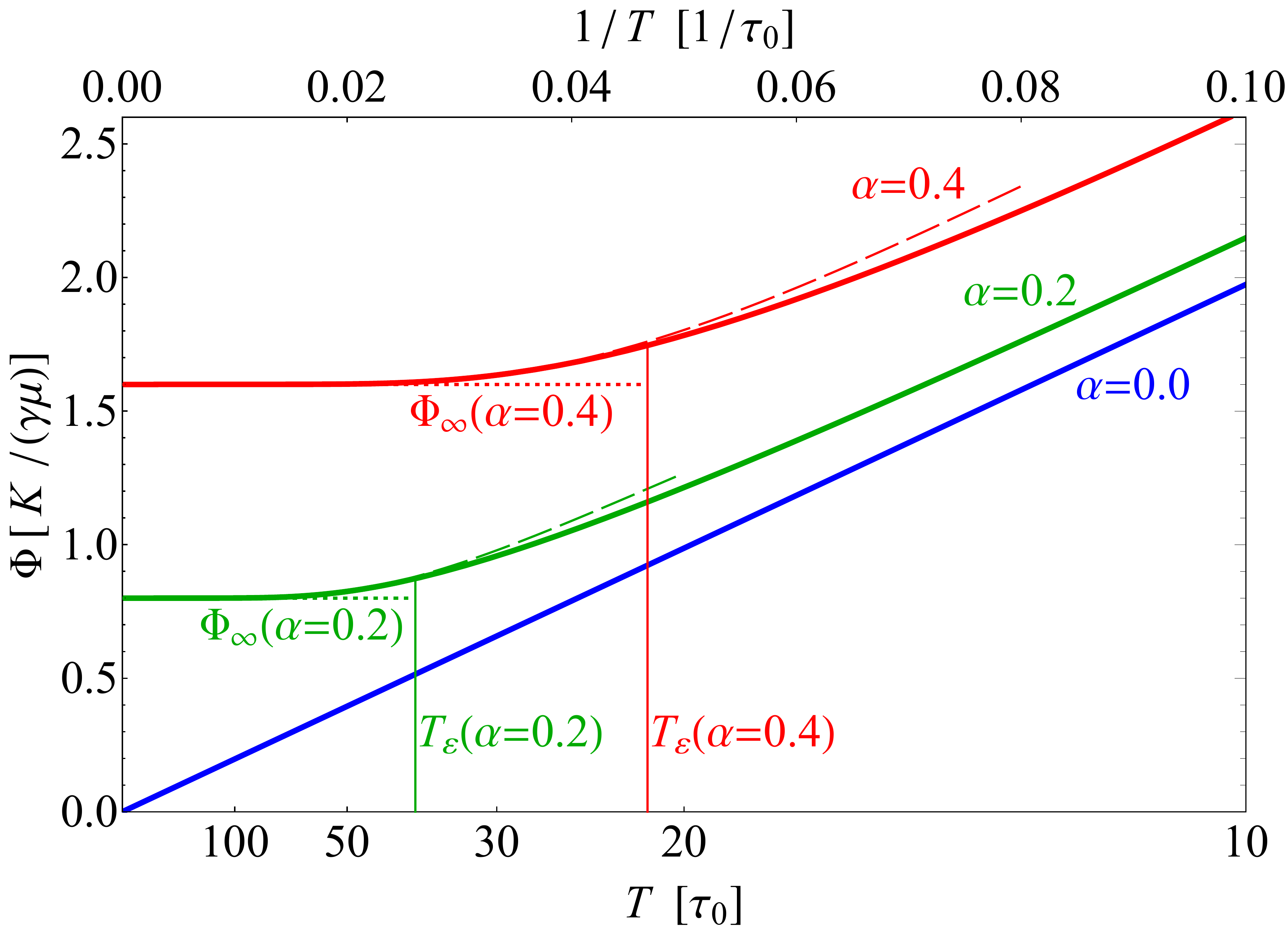}
\caption{\label{fig:cost_tinv} Minimum energy cost of magnetization switching as a function of the inverse of the switching time. 
Dashed (dotted) lines show long (infinite) switching time asymptotics. 
Thin vertical lines indicate switching time $T_\varepsilon$, for which the minimum energy cost is $\varepsilon=10\%$ larger than 
the infinite switching time limit $\Phi_\infty$.
}
\end{figure}

Substitution of Eq.~(\ref{eq:b_amp}) into Eq.~(\ref{eq:cost}) leads to the following formula for the minimum energy cost: 
\begin{equation}
    \Phi_\text{m} = \frac{2K\left[2\mathcal{E}\left(-\alpha^2 p^2\right)-\mathcal{K}\left(-\alpha^2 p^2\right)\right]}{\gamma\mu p}, \label{eq:phi}
\end{equation}
where $\mathcal{E}(.)$ is the complete elliptic integral of the second kind \cite{supplement, abramowitz1948handbook}. According to (\ref{eq:phi}), $\Phi_\text{m}$ is a monotonically decreasing (increasing) function of the switching time $T$ (damping parameter $\alpha$), as illustrated in Fig.~\ref{fig:cost_tinv}. Energy cost as a function of the switching time has two asymptotic regimes corresponding to fast and slow switching. 
For the short switching time, the magnetic potential becomes irrelevant, and the energy cost as a function of $T$ is described by a power law:
\begin{equation}
    \Phi_\text{m} \approx \frac{\pi^2 (1+\alpha^2)}{\gamma^2 T}+O(T), 
     \quad T\ll(\alpha+1/\alpha)\tau_0,\label{eq:phi_short} 
\end{equation}
The leading term in Eq.~(\ref{eq:phi_short}) specifically recovers the potential-free case. 
The power-law regime changes to an exponential dependence on $T$ for the long switching time:
\begin{equation}
    \Phi_\text{m} \approx \Phi_\infty\left(1+4\exp\left[{-\frac{\alpha T}{2\tau_0(1+\alpha^2)}}\right]\right), \quad T\gg(\alpha+1/\alpha)\tau_0, \label{eq:phi_long}
\end{equation}
which particularly demonstrates that, for a given anisotropy constant and damping parameter, the lower limit of the energy cost is $\Phi_\infty\equiv 4\alpha K/(\gamma \mu)^{-1}$, as predicted in ~\cite{barros2013microwave}. Strictly speaking, this limit is reached at infinitely long switching time, but Eq.~(\ref{eq:phi_long}) makes it possible to analyze to what extent the limit can be approached within finite $T$. In particular, termination of the reversal within time $T_\varepsilon=2\ln{(4/\varepsilon)}[\alpha+1/\alpha]\tau_0$ corresponds to the energy cost which is only by a fraction of $\varepsilon<1$ larger than $\Phi_\infty$: $\Phi_\text{m}(T_\varepsilon)/\Phi_\infty=1+\varepsilon$. Therefore, $T_\varepsilon$ has a meaning of optimal switching time in a sense that increase in $T$ beyond $T_\varepsilon$ does not lead to a significant gain in energy efficiency~(see Fig.~\ref{fig:cost_tinv}). 

\begin{figure}[!t]
\centering
\includegraphics[width=1.0\columnwidth]{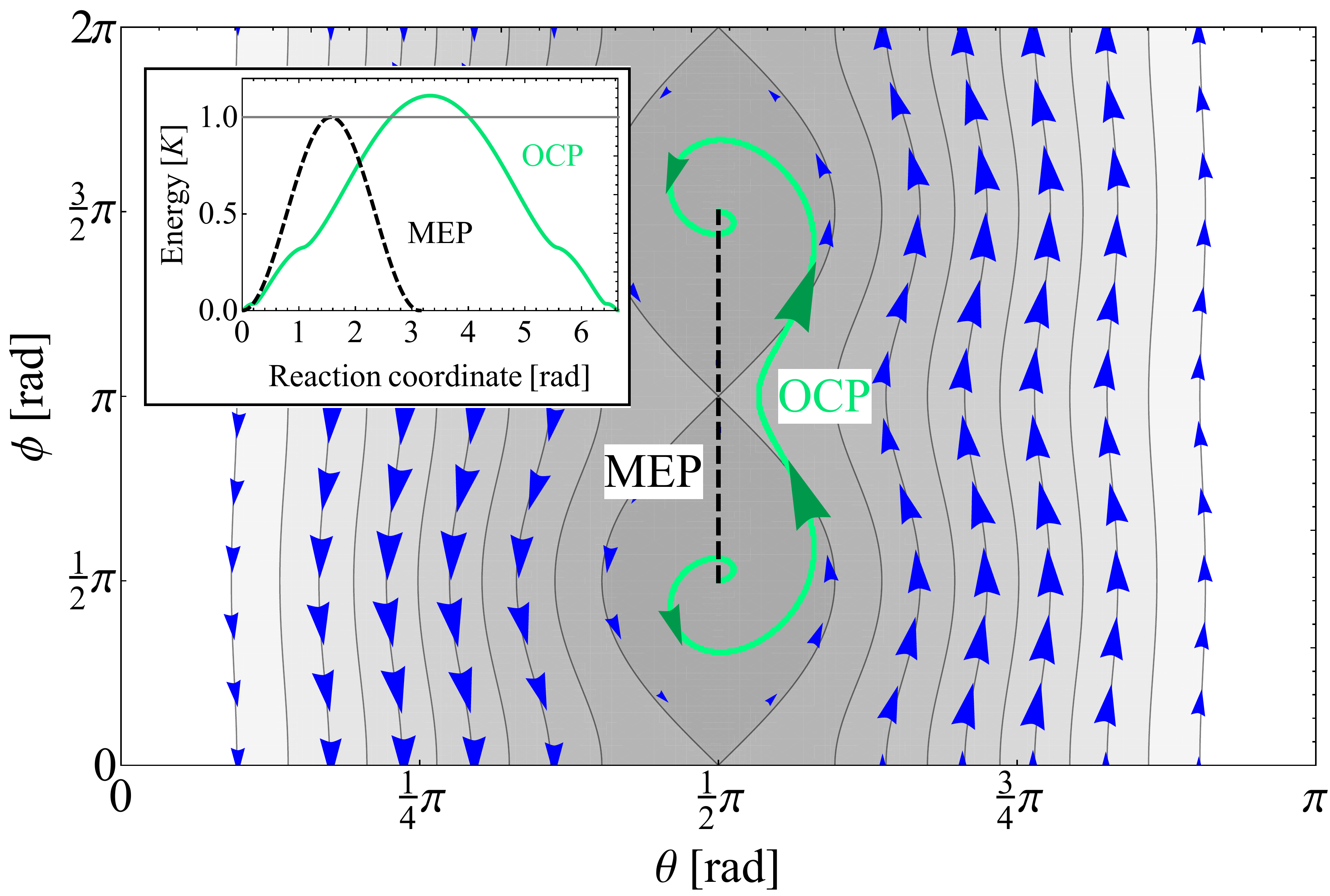}
\caption{\label{fig:2aniso} Distribution of the torque (blue arrows) generated by the internal field of the macrospin system with both easy and hard anisotropy axes, superimposed on the contour plot showing the energy surface of the system defined by Eq.~(\ref{eq:two_a}). 
Green line shows the calculated OCP for $T=0.32\tau_0$ and $\alpha=0$. Green arrows indicate the velocity of the system. The size of the blue (green) arrows code the magnitude of the torque (velocity). Black dashed line shows the minimum energy path (MEP). 
The inset shows the energy of the system as a function of displacement along the OCP (green solid line) and MEP (black dashed line). 
}
\end{figure}

Analysis of Eq.~(\ref{eq:phi}) shows that for a given switching time $T$, the energy cost is never smaller than that in a zero-potential case: $\Phi_\text{m}(T)\ge\Phi_0(T)\equiv\pi^2 (1+\alpha^2)/(\gamma^{2}T)$, where the equality is reached for $\alpha=0$. In other words, internal energy of the system can only obstruct the reversal in a system with easy-axis anisotropy, and the purpose of the switching pulse optimization is just to minimize the unfavorable effect caused by the magnetic potential in this case. To be able to use the internal energy landscape to aid the switching process,  additional terms in the magnetic potential are necessary. 
We have found that the energy cost can be reduced by adding a hard-axis anisotropy to the system. In this case, the internal energy $\tilde{E}$ can be written as:
\begin{equation}\label{eq:two_a}
    \tilde{E} = -K s_y^2 + K_{\rm h} s_z^2,
\end{equation}
where the easy axis and the hard axis are along $y$ and $z$ directions, respectively. The hard-axis anisotropy constant $K_{\rm h}$ is taken to be 10 times larger than $K$. 
The OCP between the energy minima at $s_y = \pm 1$ was obtained by a direct numerical minimization of the energy cost functional for the switching time $T=0.32\tau_0$ and damping $\alpha=0$. 
Surprisingly, the corresponding energy cost $\tilde{\Phi}_\text{m}$ turned out to be an order of magnitude smaller than that for the reversal with the same switching time and damping in the system with zero magnetic potential: $\tilde{\Phi}_\text{m}/\Phi_0\approx 0.088$. 
This phenomenon can be explained by the distribution of the internal torque, see Fig.~\ref{fig:2aniso}. Due to the hard-axis, 
there is a region in the configuration space, where the system's internal torque systematically points in the desired switching direction. By placing the switching path into this region, the optimal control efficiently exploits the internal torque to assist the switching. The external pulse has a minimal influence; its purpose is only to trigger the switching by directing the system toward the particular sector in the configuration space where the internal dynamics picks the system up and drags it to the desired target state.

Finally, we compare our OCP with another distinguished path in the configuration space -- the minimum energy path (MEP). An MEP connecting two stable states is a path lying lowermost on the energy surface, and the point of highest energy along the MEP -- a saddle point on the energy surface -- defines the energy barrier between the states, the primary quantity determining their thermal stability within harmonic rate theories~\cite{kramers_1940,vineyard_1957,brown_1979}. 
The MEP for the magnetization reversal in the system with both easy and hard axes is the shortest path connecting the energy minima through the saddle point at $\theta=\pi/2$, $\phi=\pi$ (see Fig.~\ref{fig:2aniso}). This path is very different from the calculated OCP which demonstrates a more complex structure. To emphazise the difference between MEP and OCP, we note that the OCP is a valid {\it dynamical trajectory} defined by the parameters of the equation of motion such as the switching time and damping, whereas the MEP is entirely determined by the energy surface of the system. Since the OCP does not even pass through the saddle point, the energy maximum along the OCP is higher than the energy barrier derived from the MEP (see the inset in Fig.~\ref{fig:2aniso}). This result means that optimal control of a magnetic transition which minimizes the energy spent by the external source of the switching field does not necessarily lead to a path that minimizes the energy barrier between the target states. Following an OCP involves rotation of magnetic moments in such a way that the influence of the external stimulus is minimized, but the system's internal dynamics is effectively used to aid the magnetic transition. 

In conclusion, we have presented an exact analytical solution to the problem of optimal switching of a nanomagnet. The easy-axis anisotropy alone can only increase the energy cost of the switching compared to the free-macrospin case, and unfavorable effect of the anisotropy is minimized by following the calculated OCP. The system's internal torque can be used to aid the switching by introducing a hard anisotropy axis. Our results deepen the understanding of the optimal control of magnetization switching in nanoparticles and provide guiding principles for the design of energy-efficient digital devices based on magnetic elements. 

\begin{acknowledgments}
The authors would like to thank H. J\'onsson, B. Hj\"orvarsson, V. Kapaklis and T. Sigurj\'onsd\'ottir for helpful discussions. This work was funded by the Russian Science Foundation (Grant No. 19-72-10138), the Icelandic Research Fund (Grant No. 184949-052), the Deutsche Forschungsgemeinschaft (DFG grant No. BE2464/17-1), and the Alexander von Humboldt Foundation.  
\end{acknowledgments}

%
\end{document}